\documentclass[aps,pra,preprint,showpacs,groupedaddress]{revtex4-1}

\newcommand{\beq}{\begin{equation}}
\newcommand{\eeq}{\end{equation}}
\newcommand{\beqa}{\begin{eqnarray}}
\newcommand{\eeqa}{\end{eqnarray}}
\newcommand{\ba}{\begin{array}}
\newcommand{\ea}{\end{array}}

\newcommand{\la}{\left\langle}
\newcommand{\ra}{\right\rangle}

\newcommand{\beqn}{\begin{equation*}}
\newcommand{\eeqn}{\end{equation*}}
\newcommand{\beqan}{\begin{eqnarray*}}
\newcommand{\eeqan}{\end{eqnarray*}}
\usepackage{longtable,color}
\usepackage{graphicx}
\usepackage{lineno}

\begin{document}


\title{Theoretical isotope shifts in neutral barium}


\author{C. Naz\'e}
\email[]{cnaze@ulb.ac.be}
\author{J.G. Li}
\altaffiliation{Present address: Data Center for High Energy Density Physics, Institute of Applied Physics and Computational Mathematics, PO Box 8009, Beijing 100088, China}
\author{M. Godefroid}
\email[]{mrgodef@ulb.ac.be}
\affiliation{Service de Chimie quantique et Photophysique, Universit\'e Libre de Bruxelles, 1050 Brussels, Belgium}


\date{\today}

\begin{abstract}
The present work deals with a set of problems in isotope shifts of neutral barium spectral lines.
Some well known transitions ($6s^2~^1S_0-6s6p~^{1,3}P^o_1$ and $6s^2~^1S_0-6p^2~^3P_0$) are first investigated. 
Values of the changes in the nuclear mean-square charge radius are deduced from the available experimental isotope shifts using our ab initio electronic factors. The three sets $\{ \delta\langle r^2\rangle^{A,A'}\} $ obtained from these lines are consistent with each other. 
The combination of the available nuclear mean-square radii 
with our electronic factors for the $6s5d~^3D_{1,2} -6s6p~^{1}P^o_1$ transitions produces isotope shift values in  conflict with the laser spectroscopy measurements of Dammalapati \textit{et al.} (Eur. Phys. J. D 53, 1 (2009)).
\end{abstract}


\maketitle

\section{Introduction}

The first isotope shift (IS) measurements on barium have been done by Arroe~\cite{Arr:50a}, who studied the resonance transition $6s^2~^1S_0-6s6p~^1P^o_1$ of neutral barium located at a wavelength of $\lambda_1=553.7$~nm. One of the first attempts to obtain the differences of nuclear mean-square charge radii of the radioactive barium nuclides $^{140-134}$Ba was done by Fischer \textit{et al.}~\cite{Fisetal:74a} who investigated experimentally and theoretically the isotope shift in the Ba~II resonance line $6s~^2S_{1/2}-6p~^2P^o_{1/2}$ at $\lambda =493.4$~nm. Their results are often used by other authors. 
Later, the Doppler-free spectroscopy was explored by~Nowicki \textit{et al.}~\cite{Nowetal:77a, Nowetal:78a} who studied the neutron-deficient isotopes $^{131}$Ba and $^{128}$Ba. Using the same procedure Bekk \textit{et al.}~\cite{Beketal:79a} pursued the work of Nowicki \textit{et al.} with other unstable isotopes. Concurrently but independently, Baird \textit{et al.} \cite{Baietal:79a} proposed a similar experiment.
Years later, thanks to high-resolution laser spectroscopy Grundevik \textit{et al.}~\cite{Gruetal:82a} explored the far-red transitions between the $6s5d$ and $6p5d$ configurations and, in continuity, investigated the spin-forbidden transition $6s^2~^1S_0-6s6p~^3P^o_1$ at $\lambda_2=791.3$~nm \cite{Gruetal:83b}. 
In a work of Mueller \textit{et al.}~\cite{Mueetal:83a}, collinear laser spectroscopy has been connected to the mass separator ISOLDE-II at CERN. One of the aim of this experiment was to extend the knowledge on fundamental nuclear properties into regions far from stability. In that context, isotope shifts of barium isotopes in the mass range 122-146 have been measured for the $\lambda_1$ atomic transition. 
Finally, Wijngaarden and Li~\cite{WijLi:95a} (re)measured IS of the same transition using a ring dye laser and obtained the most recent and precise IS values of this line. Many other measurements were also reported on highly excited states of Ba~I \cite{HogEli:83a,LeeHog:84a}.

The theoretical barium studies are far less advanced. In 1974, Trefftz \textit{et al.}~\cite{Tre:74a} performed calculations on various states of barium using MCHF wave functions generated with the code of Froese Fischer \cite{Fro:72a} but did not study IS. King and Wilson \cite{KinWil:85a} used a modified version of the latter program that includes the mass velocity and Darwin terms in order to calculate electron densities at the nucleus.
Besides, Fricke \textit{et al.}~\cite{Frietal:83a} and Olsson \textit{et al.}~\cite{Olsetal:88b} 
 used MCDHF wave functions to compute  electronic~$F$ factors. Finally, in an unpublished work, Kozlov and Korol \cite{KozKor:2007a} calculated the field and mass shifts of Ba~I and Ba~II using second order many-body perturbation theory to take core-valence and valence correlations into account.

Our interest has been picked up by the paper of Dammalapati \textit{et al.}~\cite{Dametal:2009a} who 
reported the first laser spectroscopy measurements of the $6s5d~^3D_{1,2} -6s6p~^{1}P^o_1$ transitions for several isotopes. 
Observing that the resulting isotope shifts strongly deviate from their expected behavior for odd isotopes in an analysis based on King-plots, the authors pointed out that there were no theoretical calculations available for comparison.

\section{Isotope shift theory}
The main ideas of the isotope shift theory are hereafter outlined. The interested reader should look at the pioneer works of Shabaev~\cite{Sha:85a,Sha:88b,Sha:88a, ShaArt:94a,Tupetal:2003a} and Palmer~\cite{Pal:88a} who expressed the theory of the relativistic mass shift as used in the present work. The tensorial form of the relativistic recoil operator was derived by Gaidamauskas \textit{et al.}~\cite{Gaietal:2011a}. Based on those developments a module, called \textsc{ris3} (Relativistic Isotope Shift) was designed \cite{Nazetal:2013a} for the revised version of\textsc{grasp2K} package~\cite{Jonetal:2013a}.

\subsection{Mass shift}
The finite mass of the nucleus gives rise to a recoil effect, called the mass shift~(MS).
The nuclear recoil corrections within the $(\alpha Z)^4m^2/M$ approximation are obtained by evaluating the expectation values of the operator
\begin{eqnarray}
\label{eq:H_MS}
   {H}_{\rm MS}  =
 \frac{1}{2M}
 \sum_{i,j} ^{N} \;
\left( {\bf  p}_{i} \cdot {\bf  p}_{j} - \frac{\alpha Z}{r_i} \left(\mbox{\boldmath $ \alpha$}_{i} + \frac{\left(\mbox{\boldmath $ \alpha$}_i\cdot{\bf  r}_i\right) {\bf  r}_i}{r^{2}_{i}}\right)\cdot  {\bf  p}_j\right).\nonumber\\
\end{eqnarray}
Separating the one-body $(i=j)$ and two-body $(i \ne j)$  terms  that respectively constitute the normal mass shift (NMS) and specific mass shift (SMS) contributions, the Hamiltonian~(\ref{eq:H_MS}) becomes
\begin{eqnarray}
\label{eq:separation}
{H}_{\rm MS} = {H}_{\rm NMS} + {H}_{\rm SMS} \; .
\end{eqnarray}
The (mass-independent) normal mass shift $K_{\rm NMS}$ and specific mass shift $K_{\rm SMS}$ parameters are defined by the following expressions
\begin{eqnarray}
\label{eq:NMS}
K_{\rm NMS} \; &\equiv&    M\langle \Psi|
   {H}_{\rm NMS}| \Psi \rangle \;,     \\
\label{eq:SMS}
K_{\rm SMS} &\equiv &
M   \langle \Psi | {H}_{\rm SMS}| \Psi \rangle \;.
\end{eqnarray}

When discussing a transition isotope shift, one needs to consider the variation of the mass parameter from one level to another.
The corresponding line  frequency isotope mass shift  is  written as the sum of the NMS and SMS contributions:
\begin{eqnarray} \label{split_freq_mass}
\delta \nu_{k,{\rm MS}}^{A,A'}  = \delta \nu_{k,{\rm NMS}}^{A,A'}+\delta \nu_{k,{\rm SMS}}^{A,A'} \;,
\end{eqnarray}
with 
\begin{eqnarray}\label{DeltaK}
\delta \nu^{A,A'}_{k,{\rm MS}}&=&\left(\frac{M'-M}{MM'}\right)\frac{\Delta K_{\rm MS}}{h}
=\left(\frac{M'-M}{MM'}\right)\Delta \widetilde K_{\rm MS}\;,
\end{eqnarray}
where  $\Delta K_{\rm MS} =  (K_{u,\rm MS} -  K_{\ell,\rm MS} )$ is the difference of the $K_{\rm MS}$ $(=K_{\rm NMS} + K_{\rm SMS} )$ parameters of the levels involved in the transition~$k$. 
For $\widetilde K$, the unit~(GHz~u) is often used in the literature. As far as the conversion factors are concerned, we use 
$\Delta K_{\rm MS}$[$m_e E_{\rm h}]= 3609.4824~\Delta \widetilde K_{\rm MS}[{\rm GHz~u}]$. 

\subsection{Field shift}
The energy shift arising from the difference in nuclear charge distributions between two isotopes  $A$ and $A'$ for levels $i= (\ell, u)$ involved in transition $k$, 
the frequency field shift~(FS) of the spectral line $k$ can be written as~\cite{Frietal:95a,Toretal:85a,Bluetal:87a}
\begin{equation} \label{Line_FS}
 \delta \nu_{k,{\rm FS}}^{A,A'} =   \frac{\delta E_{u,{\rm FS}} ^{A,A'} - \delta E_{\ell,{\rm FS}} ^{A,A'} }{h} \approx F_k  \; \delta\langle r^2\rangle^{A,A'} \; .
\end{equation}
$F_k$ is the line electronic factor
\begin{equation} \label{electronic_factor}
F_k  =  \frac{2\pi}{3h} Z \left( \frac{e^2}{4 \pi \epsilon_0} \right)  \Delta |\Psi(0)|^2_k  \; ,
\end{equation}
proportional to the change  of the total probability density at the origin
associated with the electronic transition between levels $\ell$ and $u$. 
In this approximation, the first-order frequency field shift 
 becomes
\begin{eqnarray}\label{Line_FS_sm_neg}
\delta \nu^{A,A'}_{k,{\rm FS}} &=& F_k  \; \delta\langle r^2\rangle^{A,A'}\nonumber\\&=& \frac{Z}{3 \hbar} \left( \frac{e^2}{4 \pi \epsilon_0} \right) \;  \Delta |\Psi (0)|^2_k  \; \delta\langle r^2\rangle^{A,A'} \;.
\end{eqnarray}
\subsection{The total isotope shift}
It is easy to estimate the total line frequency shift by merely  adding the mass shift~(\ref{split_freq_mass}) and field shift~(\ref{Line_FS}) or (\ref{Line_FS_sm_neg})  contributions
\begin{eqnarray} \label{split_freq_tot}
\delta \nu_k^{A,A'}&=&\overbrace{\delta \nu_{k,{\rm NMS}}^{A,A'}+\delta \nu_{k,{\rm SMS}}^{A,A'}}^{{\delta \nu_{k,{\rm MS}}^{A,A'}}}+\delta \nu_{k,{\rm FS}}^{A,A'} \\
&=&\left(\frac{M'-M}{MM'}\right)\Delta \widetilde K_{\rm MS}+F_k  \; \delta\langle r^2\rangle^{A,A'}\; \label{eq:split_freq_tot_expl}.
\end{eqnarray}

\section{Computational procedure\label{sec:Comp}}
The multiconfiguration Dirac-Hartree-Fock (MCDHF) method~\cite{Gra:2007a}, as implemented in the program package \textsc{grasp2K}~\cite{Jonetal:2007a,Jonetal:2013a},
is employed to obtain approximate wave functions that we will refer to as atomic state wave functions (ASFs). An ASF is represented by a linear combination of configuration state
functions (CSFs) with same parity $P$, total angular momentum $J$ and component $M_J$ along $z$-direction 
\begin{equation}
\label{ASF}
\Psi(\gamma PJM_J)   = \sum_{j=1}^{NCSFs} c_{j} \; \Phi(\gamma_{j}PJM_J),
\end{equation}
where $\{c_i\}$ are the mixing coefficients and $\{\gamma_i\}$ the sets of quantum numbers needed for specifying unambiguously CSFs. 
The latter are built from single-electron orbital wave functions. Applying the variational principle, the mixing coefficients and single-electron orbital wave functions are optimized simultaneously in the relativistic self-consistent field (RSCF) method.
The energy functional is estimated from the expectation value of the Dirac-Coulomb (DC) Hamiltonian~\cite{Gra:2006a}
\begin{eqnarray}
\label{eq:MCDF}
H_{DC}  =
\sum_{i=1}^N
\Big( 
c~\mbox{\boldmath $ \alpha$}_i \cdot 
{\bf  p}_i + (\beta_i-1)c^2 +V({r_i})\Big)
 + \sum_{i < j}^N
\frac{1}{r_{ij}}\;,\nonumber \\
\end{eqnarray}
where $V({r_i})$ is the monopole part of the electron-nucleus interaction, \mbox{\boldmath $ \alpha$} and $\beta$ are the $(4\times 4)$ Dirac matrices and $c$ is the speed of light ($c = 1/\alpha $ in atomic units, where $\alpha$ is the fine-structure constant). 
\begin{table*}
\begin{footnotesize}
\caption{Sample from several sources of the nuclear rms radii and difference of the nuclear mean-square charge radii relative to $^{138}$Ba, in fm and fm$^2$, respectively.\label{tab:comp_rmsBa}}
\begin{ruledtabular}
\begin{tabular}{@{}l|cccccccccccll}
\multicolumn{1}{c}{ }&\multicolumn{2}{c}{Semi-empirical formula (\ref{eq:rms_caca})}&&\multicolumn{2}{c}{From Angeli \cite{Ang:2004a}}&&\multicolumn{2}{c}{From Angeli and Marinova \cite{AngMar:2013a}}\\
\cline{2-3}\cline{5-6}\cline{8-9}
\multicolumn{1}{c}{\phantom{\bigg(}}&$\la r^2\ra^{1/2} $&$\delta\la r^2\ra^{138,A} $&&$\la r^2\ra^{1/2} $&$\delta\la r^2\ra^{138,A} $&&$\la r^2\ra^{1/2} $&$\delta\la r^2\ra^{138,A} $\\[-0.1cm]
\hline
$^{138}$Ba&4.8901& 0.0		&&4.8385(48)&0.0	&&4.8378(46) &0.0	\\
$^{137}$Ba&4.8798& 0.1006	&&4.8326(48)&0.0571&&4.8314(47) &0.0609(2)	\\
$^{136}$Ba&4.8692& 0.2040	&&4.8327(48)&0.0561&&4.8334(46) &0.0422(2)	\\
$^{135}$Ba&4.8586& 0.3071	&&4.8273(48)&0.1082&& 4.8294(47)&0.0812(3)	\\
\hline
&\multicolumn{8}{c}{$\delta\la r^2\ra^{138,A}$ values from different sources}\\
\cline{2-9}
\multicolumn{1}{c|}{ }&Ref. \cite{Nowetal:78a}&Ref. \cite{Beketal:79a}&&Ref. \cite{Baietal:79a}&Ref. \cite{Sheetal:82a}  &&Ref. \cite{Gruetal:82a}&Ref. \cite{WijLi:95a}\\
$^{138}$Ba	&0.0&0.0&&0.0&0.0&&0.0&0.0\\
$^{137}$Ba	&0.049(2)&0.059(4)\phantom{0}&&0.067&0.072(6)&&0.059(6)\phantom{0}&0.049\\
$^{136}$Ba	&0.034(3)&0.041(8)\phantom{0}&&0.061&0.044(4)&&0.042(7)\phantom{0}&0.034\\
$^{135}$Ba	&0.066(5)&0.079(12)&&0.108&0.088(7)&&0.080(10)	&0.065\\
$^{134}$Ba	&0.044(7)&0.053(16)&&0.095&0.051(7)&&0.056(10)& /
\end{tabular}\end{ruledtabular}\end{footnotesize}
\end{table*}

To effectively capture major electron correlation, CSFs of a particular parity and symmetry are generated through substitutions within an active set of orbitals occupied in the reference configuration. 
As regards of the hardware and software limitations, it is obviously impossible to include all CSF in the sense of a CAS expansion. The CSFs expansions have to be constrained so that the major correlation excitations are taken into account. In our calculations an approach based on single (S) and restricted double (rD) substitutions was applied \cite{Bieetal:2008a}. Restricted double substitutions limit the excitations to maximum one hole in the core.
In the case of barium ([Xe] $6s^2$, [Xe] $6s6p$, [Xe] $6p^2$, [Xe] $6s5d$), double excitations are applied to the valence shells but with the restriction that at most one electron is substituted from $1s2sp3spd4spd5sp$ [Xe] shells, the other(s) involving the $6s$, $5d$ or $6p$ valence shells of the considered reference configuration. 
All occupied orbitals in the reference configuration are treated as spectroscopic  and are obtained in the DHF single-configuration approximation. These orbitals are frozen in all subsequent calculations. The $J$-levels belonging to a given term were optimized simultaneously with standard weights through the EOL scheme and the set of virtual orbitals is increased layer by layer. Starting from the $n=9$ correlation layer, the core is fully opened but the new CSF are generated by single excitations from the reference.
The effect of adding the Breit interaction to the DC Hamiltonian (\ref{eq:MCDF}) is estimated to be much smaller than the  uncertainty in the transition isotope shift parameters with respect to the correlation model. This interaction has been therefore neglected.

\section{Isotope shifts of neutral barium}
\subsection{A large ``choice'' of nuclear radii}

The reliability of the FS values obtained with the \textit{ab initio} electronic $F$ factor~(\ref{electronic_factor}) is a function of the accuracy of the calculations but also of the level of confidence on the nuclear data $\delta \la r^2\ra^{A,A'}$. Table \ref{tab:comp_rmsBa} reports, nuclear root-mean-square (rms) charge radii and mean-square charge radii differences  from several sources, taking $^{138}$Ba as the reference isotope. The aim of this table is to illustrate the difficulties to reduce uncertainties on the nuclear rms charge radii. The first set of columns is obtained with the semi-empirical formula \beqa
\label{eq:rms_caca}
R_{\rm rms} = \la r^2\ra^{1/2}=0.836A^{1/3} + 0.570 \text{ fm  if }A>9,~~~
\eeqa
and is compared with the values compiled by Angeli~\cite{Ang:2004a} and Angeli and Marinova~\cite{AngMar:2013a}. The positive sign of $\delta\la r^2\ra$ (according to the convention
$ \delta\langle r^2\rangle^{A,A'}  =   \langle r^2\rangle^{A} -  \langle r^2\rangle^{A'}$
with $A  > A'$) indicates that the neutron deficient isotopes of barium have smaller rms radii than~$^{138}$Ba.
\begin{table*}
\begin{footnotesize}
\caption{Experimental measurements of isotope shifts for the transitions $6s^2~^1S_0-6s6p~^{1}P^o_1$ and $6s^2~^1S_0-6s6p~^3P^o_1$ (in~GHz). \label{tab:barium1P1_1S0}}
\begin{ruledtabular}
\begin{tabular}{lllllll}
&&138&\multicolumn{1}{c}{137}&\multicolumn{1}{c}{136}&\multicolumn{1}{c}{135}&\multicolumn{1}{c}{134}\\
\hline
\multicolumn{2}{l}{$6s^2~^1S_0-6s6p~^{1}P^o_1$}\\
&Ref. \cite{WijLi:95a} (1995)&0.0&$-0.21515(16)$	&$-0.12802(39)$	&$-0.25929(17)$	&\multicolumn{1}{c}{/}\\
&Ref. \cite{Beketal:79a} (1979)	&0.0&$-0.2150(7)$	&$-0.1289(5)$	&$-0.2609(7)$	&$-0.1430(5)$	\\
&Ref. \cite{Baietal:79a} (1979)	&0.0&$-0.2147(5)$	&$-0.1275(13)$&$-0.2587(7)$	&$-0.1428(12)$\\
&Ref. \cite{Arr:50a}	(1950)	&0.0&$-0.16(2)$	&$-0.066(20)$	&$-0.22(2)$	&$-0.13(2)$\\
\\
\multicolumn{2}{l}{$6s^2~^1S_0-6s6p~^3P^o_1$}\\
&Ref. \cite{Gruetal:83b} (1983)	&0.0&$-0.1834(10)$	&$-0.1092(10)$&$-0.2199(10)$&$-0.1223(25)$\\
\end{tabular} \end{ruledtabular}\end{footnotesize}
\end{table*}

The second-half of the table reports the $\delta\langle r^2\rangle^{138,A}$ values deduced from different techniques. 
Bekk  \textit{et al.} \cite{Beketal:79a} measured IS by laser induced resonance fluorescence on an atomic beam of a sample of barium. In continuity of the work of Nowicki \textit{et al.}~\cite{Nowetal:78a}, they connected their IS measurement of the $^1S_0-~^1P^o_1$ transition in Ba~I to the data of Fischer \textit{et al.}~\cite{Fisetal:74a} on the $\lambda=493.4$~nm transition of Ba II via a King plot. 
Thanks to the electronic factor $F$ 
and SMS 
value deduced in~\cite{Fisetal:74a},  
Bekk  \textit{et al.} \cite{Beketal:79a} were able to get their own $F$ factor. They deduced the NMS through the relation
\begin{eqnarray}\label{eq:NMS_expe}
 \delta \nu_{\rm NMS}^{A,A'} \simeq\left(\frac{m_e}{M'}-\frac{m_e}{M}\right) \nu_{exp}\;.
\end{eqnarray}
and, assuming that $\delta\nu_{\rm SMS}=(0\pm1)\delta\nu_{\rm NMS}$~\cite{Nowetal:78a},  obtained the nuclear mean-square charge radii  values. As pointed out by the authors themselves, the latter approximation is the major source of the uncertainties on $\delta\langle r^2\rangle$.
 Baird \textit{et al.}~\cite{Baietal:79a} combined their own results with other optical data and electronic x-ray measurements.
 Muonic x-ray measurements of the nuclear charge radii and $\delta\langle r^2\rangle$ values were presented by Shera \textit{et al.} \cite{Sheetal:82a}. By comparing IS data on the $^1S_0-~^1P^o_1$ transition with the muonic $\delta\langle r^2\rangle$ values, Shera \textit{et al.} extracted the electronic factor.  Grundevik \textit{et al.} \cite{Gruetal:82a} reevaluated the SMS values in the Ba~I and Ba~II resonance transitions and obtained results strongly consistent with \cite{Beketal:79a}.
 On their side, van Wijngaarden and Li~\cite{WijLi:95a} used the procedure proposed by~\cite{Nowetal:78a} on their new IS measurements. Using the electronic factor of Fischer \textit{et al.}~\cite{Fisetal:74a}, they deduce a new value of the variation of the nuclear mean-square $\delta\la r^2\ra$, in very good agreement with~\cite{Nowetal:78a}.

It is worthwhile to notice that in some cases the rms charge radius $\la r^2 \ra^{1/2}$ decreases when the number of neutrons increases along the isotope chain. This  will never be reflected when using the semi-empirical formula~(\ref{eq:rms_caca}) whose values increase monotonically. Therefore, in present paper, this approach will be left aside when rms charge radii are needed, in favor of data coming from the literature such as Angeli (and Marinova)'s compilations~\cite{Ang:2004a,AngMar:2013a}.
In addition, table~\ref{tab:comp_rmsBa} demonstrates the difficulties to isolate the nuclear rms radius and sheds light on the remaining large uncertainties on the $\delta\langle r^2\rangle$ nuclear data for these systems. 

\subsection{Some well-known transitions}
The resonance transition $6s^2~^1S_0-6s6p~^{1}P^o_1$ is maybe the most well-known in barium and is, together with the intercombination line $6s^2~^1S_0-6s6p~^3P^o_1$, a good starting point of our analysis.$~^{138}$Ba is the most abundant isotope of barium on Earth with its 82 neutrons and as such is often chosen as the pivot; the IS relative to the isotope 138 are given from $A=137$ to 134 in table \ref{tab:barium1P1_1S0}. Some details behind those measurements are given in the introduction of the present paper. 
The values of Arroe~\cite{Arr:50a}, originally given in cm$^{-1}$, were converted so that the given error bars of 
$\pm 0.7\times10^{-3}$cm$^{-1}$ become $\pm21$ MHz.

The negative signs indicate that the isotope $^{138}$Ba has the lowest line frequency for each $^{138,A'}$Ba isotope pair ($A'<138$). 
Assuming the dominance of the FS, these isotopes behave in the most current way, considering the density reduction with electronic excitations (i.e. $\delta\nu_{\rm FS}^{A,A'}<0$ with $A>A'$).
The consistency between experiments is rather good.

In Neugart \textit{et al.}~\cite{Neuetal:81a} and Mueller \textit{et al.}~\cite{Mueetal:83a}, IS measurements are reported for neuron-rich isotopes $^{139}$Ba to $^{144}$Ba and the mean-square nuclear charge radii $\delta\langle r^2\rangle$ have been estimated, following the procedure of Bekk \textit{et al.}~\cite{Beketal:79a} in which the unknown SMS is taken to be of the order of the normal mass shift. These results are presented in the separated table~\ref{tab:Nrich}.
The relativistic coupled-cluster approach has been used by M\aa rtensson-Pendrill and Ynnerman~\cite{MarYnn:92a} to calculate an electronic $F$ factor that allowed the authors to revise values of nuclear charge radii using optical isotope shifts for $^{122-148}$Ba and muonic results for the stable isotopes $^{114-138}$Ba. Muonic results $\delta\langle r^2\rangle_\mu$ were used to derive a $K_{\rm SMS}$ parameter, itself used to extract $\delta\langle r^2\rangle_{\rm opt}$. The error bars reflect the uncertainty in the SMS,  in $F$ and in the optical data. The agreement within the three sets of of $\delta\langle r^2\rangle$ is quite satisfactory. 

\begin{table*} 
\caption{Experimental measurements of neutron-rich isotope shifts for the transitions $6s^2~^1S_0-6s6p~^{1}P^o_1$ (in GHz) and their related difference nuclear mean-square radii with the reference isotope $^{138}$Ba. \label{tab:Nrich}}
\begin{ruledtabular}
\begin{tabular}{lllll}
&&\multicolumn{1}{c}{141}&\multicolumn{1}{c}{140}&\multicolumn{1}{c}{139}\\
\hline
\multicolumn{2}{l}{$6s^2~^1S_0-6s6p~^{1}P^o_1$}\\
\multicolumn{2}{l}{Ref.~\cite{Neuetal:81a}}&$\phantom{-}1.505(8)$&$\phantom{-}1.075(6)$&$\phantom{-}0.473(6)$\\
\multicolumn{2}{l}{Ref.~\cite{Mueetal:83a}}&$\phantom{-}1.505(5)$&$\phantom{-}1.075(3)$&$\phantom{-}0.473(3)$\\
\\
&&$\delta \langle r^2\rangle^{138,141}$&$\delta \langle r^2\rangle^{138,140}$ &$\delta \langle r^2\rangle^{138,139}$\\\cline{3-5}\\[-.4cm]
\multicolumn{2}{l}{Ref.~\cite{Neuetal:81a} (1981)}&$-0.395(13)$&$-0.281(9)$&$-0.124(5)$\\
\multicolumn{2}{l}{Ref.~\cite{Mueetal:83a} (1983)}&$-0.395$&$-0.281$&$-0.124$ \\
\multicolumn{2}{l}{Ref.~\cite{MarYnn:92a} (1992)}&$-0.440(1)(13)(1)$&$-0.314(1)(9)(1)$&$-0.1381(5)(41)(9)$\\
\multicolumn{2}{l}{Ref. \cite{AngMar:2013a} (2013)}&$-0.410$&$-0.292$&$-0.129$\\
\end{tabular}\end{ruledtabular} \end{table*}

\subsubsection{Mass shift calculation}
The results of our calculations for the mass factors of transitions $6s^2~^1S_0-6s6p~^{1}P^o_1$ and $6s^2~^1S_0-6s6p~^3P^o_1$ are reported in table~\ref{tab:K_MS1}. For each transition, the first column gives the value of the $\widetilde K_{\rm MS}$ parameters in GHz~u, the second column is the value of the MS in MHz for the isotopic pair $^{138,136}$Ba (i.e. multiplying the $\widetilde K_{\rm MS}$ parameter by the mass factor ($1/M_{138}-1/M_{136}$)). Nuclear masses  $(M)$ are calculated by substracting the mass of the electrons and the binding energy from the atomic mass $(M_{\rm atom})$, using the formula: 
\begin{eqnarray}\label{nu_mass}
M(A,Z) = M_{\rm atom}(A,Z)-Zm_e +B_{el}(Z)
\end{eqnarray}
where the total electronic binding energy (expressed in eV) is estimated using \cite{Huaetal:76a, Lunetal:2003a}
\begin{eqnarray}\label{Bind_ell}
B_{el}(Z)=14.4381~Z^{2.39} + 1.55468\cdot10^{-6}~Z^{5.35} \; .
\end{eqnarray}
 Atomic masses are provided by~\cite{NISTAtW2}.
\begin{table}[ht!]
\caption{Values of the $\Delta\widetilde K_{\rm MS}$ parameters for the $6s^2~^1S_0-6s6p~^{1}P^o_1$ and $6s^2~^1S_0-6s6p~^3P^o_1$ transitions (in GHz u) and values of the MS (in MHz) for the $^{138,136}$Ba pair. \label{tab:K_MS1}}\begin{ruledtabular}\begin{tabular}{l|rrrrrrrrr}
&&\multicolumn{3}{c}{$6s^2~^1S_0-6s6p~^{1}P^o_1$}&&\multicolumn{3}{c}{$6s^2~^1S_0-6s6p~^3P^o_1$}\\
&&\multicolumn{1}{c}{$\Delta\widetilde K_{\rm MS}$}&&\multicolumn{1}{c}{MS}
 &&\multicolumn{1}{c}{$\Delta\widetilde K_{\rm MS}$}&&\multicolumn{1}{c}{MS}\\
\cline{3-5}\cline{7-9}
DHF	&&296.65	&$\rightarrow$&$-31.68$		&&242.69		&$\rightarrow$&$-25.92$&\\
6SrD	 	&&$-393.26$	&$\rightarrow$&42.00	&&$-125.10$	&$\rightarrow$&13.36&\\
7SrD	 	&&$-21.81$	&$\rightarrow$&2.33		&&$-203.07$	&$\rightarrow$&21.69&\\
8SrD 	&&$37.28$	&$\rightarrow$&$-3.98$	&&$-183.89$	&$\rightarrow$&19.64&\\
9S	  	&&$61.66$	&$\rightarrow$&$-6.58$	&&$-184.19$	&$\rightarrow$&19.67&\\
10S 		&&$-59.65$	&$\rightarrow$&6.37		&&$-180.10$	&$\rightarrow$&19.23&\\
11S		&&$-61.22$	&$\rightarrow$&6.54		&&$-179.14$	&$\rightarrow$&19.13&\\
\end{tabular}\end{ruledtabular}\end{table}

Clearly, correlation effects play a major role and core excitations are non-negligible. However, a comparison of the MS with the isotope shift values of the table~\ref{tab:barium1P1_1S0} reveals that the contribution of MS is really small and represents only 5\% of the total isotope shift value for the resonance transition but reaches  around 20\% for the transition $^1S_0-~^3P^o_1$. In consequence, for both lines one expects a much larger contribution of the FS. Furthermore, one observes a much larger stability and better convergence of the transition parameter $\Delta\widetilde K_{\rm MS}$ for the spin-forbidden transition (with a 0.5\% difference between the $10S$ and $11S$ values) than for the resonance transition suffering from oscillations, even in the largest calculations. Actually, the values of $\widetilde K_{\rm MS}$ of the $^1P^o_1$ and $^1S_0$ levels are so close that the slightest change on the level parameter strongly affects the transition parameter $\Delta\widetilde K_{\rm MS}$. 
The accuracy of the latter is therefore hard to evaluate but its reliability is discussed further in relation with the $\delta\langle r^2\rangle$ values that can be deduced from the experimental IS (see table~\ref{tab:ourR2}).
Behind the value of the mass parameter $\Delta\widetilde K_{\rm MS}=-61.22$~GHz~u for the transition $\lambda_1$, hides the sum of the $\Delta\widetilde K_{\rm NMS}=362.51$~GHz~u and $\Delta\widetilde K_{\rm SMS}=-423.74$~GHz~u. By looking at them, it seems that the approximation $\delta\nu_{\rm SMS}=(0\pm1)\delta\nu_{\rm NMS}$ proposed by Nowicki \textit{et al.}~\cite{Nowetal:78a} is not senseless.

\subsubsection{Field shift calculation}

The level field shift in Ba~I is around $10^{-4}E_{\rm h}$ for both states $6s^2~^1S_0$ and $6s6p~^{1}P^o_1$, while the transition FS is $10^{-8}E_{\rm h}$; a good accuracy is not easy to reach especially for a total binding energy around $-8000E_{\rm h}$. With that respect, the formalism~(\ref{Line_FS_sm_neg}) is more reliable in view of the extreme difficulty to obtain highly converged total energies.
Furthermore, the perturbative approach offers the freedom to explore and seek for the best nuclear mean-square radius. 

Table \ref{biblio_F} gives a chronological list of the experiments and calculations performed so far to determine the electronic $F$ factor of the $6s^2~^1S_0-6s6p~ ^1P^o_1$ and $6s^2~^1S_0-6s6p~ ^3P^o_1$ transitions. In the second-half of the table are presented the results of our calculations, starting from Dirac-Hartree-Fock to our most correlated model. 
\begin{table}
\caption{Comparison of different theoretical and experimental determinations of the electronic $F$ factor.\label{biblio_F}}\begin{ruledtabular}
\begin{tabular}{@{}lrccc}
&&\multicolumn{2}{c}{$F$ (GHz/fm$^2$)}\\
\cline{3-4}
&&$6s^2~^1S_0-6s6p~ ^1P^o_1$
&$6s^2~^1S_0-6s6p~ ^3P^o_1$\\
\hline\multicolumn{2}{l}{Theory}\\
\multicolumn{2}{r}{Ref. \cite{Gruetal:83b}} 	&$-2.34\phantom{(26)}$&$-2.34\phantom{(26)}$\\				
\multicolumn{2}{r}{Ref. \cite{Frietal:83a}} 		&$-2.99\phantom{(26)}$&$-2.55\phantom{(26)}$\\
\multicolumn{2}{r}{Ref. \cite{Toretal:85a}}	&$-2.996\phantom{(6)}$&$-2.546\phantom{(6)}$\\
\multicolumn{2}{r}{Ref. \cite{KinWil:85a}}\\
\multicolumn{2}{r}{ Ref. \cite{Olsetal:88b}}	&$-2.998\phantom{(6)}$&$-2.544\phantom{(6)}$\\
\multicolumn{2}{r}{Ref. \cite{MarYnn:92a}} 	&$-3.39(10)$\\\\
\multicolumn{2}{l}{Experiment}\\
\multicolumn{2}{r}{Ref. \cite{Beketal:79a}} 	&$-3.929\phantom{(2)}$\\				
\multicolumn{2}{r}{Ref. \cite{Sheetal:82a}}  	&$-3.04(26)$\\
\multicolumn{2}{r}{Ref. \cite{Frietal:83a}}		&					&$-2.59(22)$\\
\multicolumn{2}{r}{Ref. \cite{Kunetal:83a}} 	&$-3.99(65)$					\\\\
\multicolumn{2}{l}{This work}	&\multicolumn{2}{c}{\textsc{grasp2K}}\\
\cline{3-4}
&(DHF)							& $-3.48$&					$-2.76$\\
&\vdots&\vdots&\vdots\\
&(10S)                           & $-3.95$					&$-3.49$\\
&(11S)							& $-3.95$ 						&$-3.49$\\
\end{tabular}
\end{ruledtabular}\end{table}

To our knowledge, it is the first time that such a highly correlated model is used on barium. This could be an explanation for the fact that our values present some differences with the literature. 
For the $^1S_0-~^1P_1^o$ transition, our values are in favor of a large $F$ factor, in line with Bekk \textit{et al.} \cite{Beketal:79a} and Kunold \textit{et al.}~\cite{Kunetal:83a}. For the spin-forbidden line, our $F$ factor values are larger (13\%) than any others available in the literature. 
The field-shift parameters presented in table \ref{biblio_F} have converged within 0.1\% for both transitions.

\subsubsection{$\delta\langle r^2\rangle^{A,A'}$ extraction}
Thanks to the $\Delta\widetilde K_{\rm MS}$ parameters of table~\ref{tab:K_MS1}, the electronic $F$ factor of table~\ref{biblio_F} and the formula~(\ref{eq:split_freq_tot_expl}), we propose to isolate the values of $\delta\langle r^2\rangle$ that would reproduce the total measured isotope shift value (see tables \ref{tab:barium1P1_1S0} and \ref{tab:Nrich}). For the transition $6s^2~^1S_0-6s6p~^1P^o_1$, one uses the IS values of van Wijngaarden and Li that appear to be the most precise experimental values. It is possible to double-check the consistency of our results by extracting the $\delta\langle r^2\rangle$ values from the experimental IS of the other transition, i.e. $6s^2~^1S_0-6s6p~^3P^o_1$ \cite{Gruetal:82a}. These results are presented in table~\ref{tab:ourR2} and look really promising. 
Error bars on $\delta\langle r^2\rangle^{138,A}$ reflect the uncertainty on the IS measurements. The $\delta\langle r^2\rangle^{138,137}$ obtained from these two transitions differ by less than $1\%$. The $\delta\langle r^2\rangle^{138,136}$ and $\delta\langle r^2\rangle^{138,135}$ values resulting from the two lines agree within 8 to 5\%, respectively. However, the $\delta\langle r^2\rangle^{138,134}$ values reveal a discrepancy of 17\%. 

\begin{table}[h]
\caption{Extraction of $\delta\langle r^2\rangle$ values (in fm$^2$) using experimental IS values (in MHz) from the literature and our electronic factors MS (in MHz) and $F$ (in GHz fm$^2$): $\delta\langle r^2\rangle$=(IS$-$MS)/$F$. \label{tab:ourR2}} 
\begin{ruledtabular}\centering\begin{tabular}{l| ccrrrc}
\multicolumn{3}{c}{$6s^2~^1S_0-6s6p~^1P^o_1$}\\
\multicolumn{1}{c}{ }&&\multicolumn{1}{l}{(~~~IS$_{\rm exp}$~~~~~~~~$-$}&\multicolumn{1}{c}{MS~)}&\multicolumn{1}{c}{$/F$}&$\rightarrow$& $\delta\langle r^2\rangle^{138,A}$\\
\cline{3-7}
$^{141}$Ba&&$1505(8)\phantom{.000}$			&$-9.48$			&$-3.95$& $\rightarrow$ &$-0.3832(20)$\\
$^{140}$Ba&&$1075(6)\phantom{.000}$			&$-6.37$			&$-3.95$& $\rightarrow$ &$-0.2736(15)$\\
$^{139}$Ba&&$\phantom{-}473(6)\phantom{.000}$	&$-3.21$			&$-3.95$& $\rightarrow$ &$-0.1205(15)$\\
$^{137}$Ba&&$-215.15(16)$					&$\phantom{-}3.24$&$-3.95$	&$\rightarrow$&$\phantom{-}0.0553(0)\phantom{0}$\\
$^{136}$Ba&&$-128.02(39)$					&$\phantom{-}6.54$&$-3.95$	&$\rightarrow$&$\phantom{-}0.0341(1)\phantom{0}$\\
$^{135}$Ba&&$-259.29(17)$					&$\phantom{-}9.88$&$-3.95$	&$\rightarrow$&$\phantom{-}0.0681(0)\phantom{0}$\\
$^{134}$Ba&&$-143.0(5)\phantom{10}$			&13.3\phantom{8}&$-3.95$&$\rightarrow$&$\phantom{-}0.0395(13)$\\
\multicolumn{3}{c}{$6s^2~^1S_0-6s6p~^3P^o_1$}\\
&&\multicolumn{1}{l}{(~~~IS~~~~~~~~~~$-$}&\multicolumn{1}{c}{MS~)}&\multicolumn{1}{c}{$/F$}&$\rightarrow$& $\delta\langle r^2\rangle^{138,A}$\\
\cline{3-7}
$^{137}$Ba&&$-183.4(1.0)$&9.49 &$-3.49$		&$\rightarrow$&0.0553(3)\phantom{1}\\
$^{136}$Ba&&$-109.2(1.0)$&19.13&$-3.49$		&$\rightarrow$&0.0368(3)\phantom{1}\\
$^{135}$Ba&&$-219.9(1.0)$&28.90&$-3.49$		&$\rightarrow$&0.0714(3)\phantom{1}\\
$^{134}$Ba&&$-122.3(2.5)$&38.83&$-3.49$		&$\rightarrow$&0.0462(7)\phantom{1}\\
\hline\hline
\multicolumn{3}{c}{$6s^2~^1S_0-6p^2~^3P_0$}\\
&&\multicolumn{1}{l}{(~~~IS~~~~~~~~~~$-$}&\multicolumn{1}{c}{MS~)}&\multicolumn{1}{c}{$/F$}&$\rightarrow$& $\delta\langle r^2\rangle^{138,A}$\\
\cline{3-7}
$^{137}$Ba&&$-331.7(5.0)$& 21.49&$-7.09$		&$\rightarrow$&0.0498(7)\phantom{1}\\
$^{136}$Ba&&$-199.0(3.0)$& 43.33&$-7.09$		&$\rightarrow$&0.0342(4)\phantom{1}\\
$^{135}$Ba&&$-396.1(5.9)$& 65.45&$-7.09$		&$\rightarrow$&0.0651(8)\phantom{1}\\
$^{134}$Ba&&$-219.0(9.9)$& 88.94&$-7.09$		&$\rightarrow$&0.0433(14)\\
\end{tabular} \end{ruledtabular}\end{table}

The isotope shifts of the  $6s^2~^1S_0-6p^2~^3P_0$  transition reported in table~\ref{tab:ourR2}, are taken from the work of Jitschin and Meisel \cite{JitMei:80a}. They resolved the IS for several highly excited states, using Doppler-free two-photon laser spectroscopy. They needed the relevant  electronic $F$ factors to extract the $\delta\langle r^2\rangle$ values and the only response was from Olsson \textit{et al.}~\cite{Olsetal:88b}. Based on the correlation model detailed in section \ref{sec:Comp}, we calculated the IS parameters of the $6p^2~^3P_0$ state.  
The errors bars are  sensitively larger than for the two other experiments.
However, the consistency between the three independent sets give us confidence in the reliability of our electronic parameters ($K_{\rm MS}$ and $F$), which is the original point of this section.

Let us refer to the table~\ref{tab:comp_rmsBa} in order to compare these newly extracted values with the available ones. The  $\delta\langle r^2\rangle$ values from the semi-empirical formula as well as results of Baird \textit{et al.} \cite{Baietal:79a} are both out of range but there are some nice agreements with the other experiments. The results of Bekk \textit{et al.} \cite{Beketal:79a} and Grundevik \textit{et al.} \cite{Gruetal:82a} seem to be confirmed, especially for $\delta\langle r^2\rangle^{138,137}\simeq0.055$ fm$^2$ but not for $\delta\langle r^2\rangle^{138,136}$, for which, our results are closer to van Wijngaarden and Li's value~\cite{WijLi:95a} ($\delta\langle r^2\rangle^{138,136}=0.034$ fm$^2$). The $\delta\langle r^2\rangle^{138,135}$ is more disputable, but let us  just underline that our result is closer to van Wijngaarden and Li as well as the value obtained by M\aa{}rtensson-Pendrill and Ynnerman $\delta\langle r^2\rangle^{138,135}=0.0728 (15)(22)(2)$ fm$^2$~\cite{MarYnn:92a}. The discrepancy found for $\delta\langle r^2\rangle^{138,134}$ forbids us to draw any further conclusions. 
All our results for neutron-rich isotopes confirm the deductions of Neugart \textit{et al.}~\cite{Neuetal:81a} and Mueller \textit{et al.} \cite{Mueetal:83a} within 3\% (see table~\ref{tab:Nrich}). 

\section{The issues of the $^1P^o_1-~^3D_{1,2}$ transitions}

The level scheme of barium also exhibits low lying $6s5d~^1D_2,~^3D_1$ and $^3D_2$  metastable states. However, the transitions $6s5d~^3D_1-6s6p~^1P^o_1$ and $6s5d~^3D_2-6s6p~^1P^o_1$ are far less studied and the first measurements of their IS have been performed by Dammalapati \textit{et al.}~\cite{Dametal:2009a}.
The latter are displayed in table \ref{tab:valeur_Dam} using the $^{138}$Ba as the pivot. 
These authors also measured shifts of other pairs, reporting $\delta\nu^{136,134}_{\rm IS}=-84.8(8)$~MHz and $-80(1)$ MHz for the transitions $^1P^o_1-~^3D_1$ and $^1P^o_1-~^3D_2$, respectively, and $\delta\nu^{137,135}_{\rm IS}=-75.3(5)$ MHz for the resonance line frequency. 

Following the computational procedure described in the section \ref{sec:Comp}, the $F$ and $\widetilde K_{\rm MS}$ parameters have been calculated and are reported  in table~\ref{tab:K_MS2}.
\begin{table}[ht!]
\caption{Values of the electronic $F$ factor (in GHz/fm$^2$) and of the $\Delta\widetilde K_{\rm MS}$ parameters (in GHz~u) for the $6s5d~^3D_1-6s6p~^1P^o_1$ and $6s5d~^3D_2-6s6p~^1P^o_1$ transitions. \label{tab:K_MS2}}
\begin{ruledtabular}
\begin{tabular}{lccccc}
&\multicolumn{2}{c}{$F$ (GHz/fm$^2$)}&& \multicolumn{2}{c}{$\Delta\widetilde K_{\rm MS}$ (GHz~u)}\\
 \cline{2-3} \cline{5-6}
&$~^3\!D_1-~^1\!P^o_1$ &$~^3\!D_2-~^1\!P^o_1$&&
$~^3\!D_1-~^1\!P^o_1$ &
$~^3\!D_2-~^1\!P^o_1$\\
\hline
(DHF)&$0.31$&	$0.31$&&$937.12$	&$962.56$\\
\multicolumn{1}{c}{\vdots}&\multicolumn{1}{c}{\vdots}&
\multicolumn{1}{c}{\vdots}&&\multicolumn{1}{c}{\vdots}&
\multicolumn{1}{c}{\vdots}\\
(10S)&$0.75$&$0.74$&&$-284.34$	&$-324.09$\\
(11S)&$0.76$&	$0.74$&&$-311.99$	&$-294.94$\\
\end{tabular}
\end{ruledtabular}
\end{table}

It firstly appears that the electronic $F$ factor is much smaller and the $\Delta\widetilde K_{\rm MS}$ parameter larger than for the three transitions considered above (see table~\ref{tab:K_MS1}). The mass shift value is $\delta\nu_{\rm MS}^{138,136}=33.32$ MHz while the frequency FS is $\delta\nu_{\rm FS}^{138,136}=25.73$ MHz (using $\delta\langle r^2\rangle^{138,136}$=0.034 fm$^2$). Surprisingly, they are both of the same order of magnitude. The observation  that FS strongly dominates the MS in high-$Z$ systems cannot be obviously applied to neutral barium. Moreover, we would recommend the greatest  caution about the transition considered. 
Furthermore, the difference between the two most elaborated calculations of the values presented in table \ref{tab:K_MS2} is around 9\% for the two $\Delta\widetilde K_{\rm MS}$ parameters and 0.5\% for both electronic $F$ factors. Thanks to those two observations, we could estimate errors bars of 5\% on the IS values obtained from our electronic factors calculations for the $6s5d~^3D_1-6s6p~^1P^o_1$ and $6s5d~^3D_2-6s6p~^1P^o_1$ transitions.
  
Table \ref{tab:valeur_Dam} compiles the results of Dammalapati \textit{et al.} with our predictions using the $\delta \langle r^2 \rangle^{138,A}$ of other authors. The underlined data are obtained with the $\delta\langle r^2\rangle$ values that look the most reliable on the basis of  the $^{3,1}P^o_1-~^1S_0$ transitions.
\begin{table*}[ht!]\centering
\caption{Experimental measurements of isotope shifts for the transitions $6s5d~^3D_1-6s6p~^1P^o_1$ and $6s5d~^3D_2-6s6p~^1P^o_1$ (in MHz) from \cite{Dametal:2009a} compared with the IS calculated with the mean-square radii differences available in the literature and our electronic factors. \label{tab:valeur_Dam}}
\begin{ruledtabular}
\begin{tabular}{lrrrrrr}
&&138&\multicolumn{1}{c}{137}&\multicolumn{1}{c}{136}&135&\multicolumn{1}{c}{134}\\
\hline
\multicolumn{2}{l}{$6s5d~^3D_1-6s6p~^1P^o_1$}\\
\multicolumn{2}{r}{Dammalapati \textit{et al.}~\cite{Dametal:2009a}}	&0.0&$114(4)$	&$-59.3(6)$&39(4)&$-144.1(10)$ \\\\
\multicolumn{4}{l}{With the $\delta\langle r^2\rangle$ of:}\\
&Baird \textit{et al.}	\cite{Baietal:79a} 		&0.0&$67.2$		&$\phantom{-}79.5\phantom{(0)}$&$132.1$&$139.5\phantom{(0)}$\\
&Bekk \textit{et al.}~\cite{Beketal:79a}&&\underline{$61.2$}&$\phantom{-}64.4\phantom{(0)}$&$	110.1$&$107.7\phantom{(1)}$\\
&Shera \textit{et al.}~\cite{Sheetal:82a}		&&$71.0$			&$66.6\phantom{(0)}$			&$\phantom{-}116.9$&$106.2\phantom{(1)}$\\
&Grundevik \textit{et al.}~\cite{Gruetal:82a}		&&\underline{$61.2$}&$\phantom{-}65.1\phantom{(0)}$&$	110.9$&$110.0\phantom{(1)}$\\
&van Wijngaarden \textit{et al.}~\cite{WijLi:95a}	&&$53.6$&\phantom{-}\underline{$59.1$}\phantom{(0)}&$	99.5$\\\\
\hline
\multicolumn{2}{l}{$6s5d~^3D_2-6s6p~^1P^o_1$}\\
\multicolumn{2}{r}{Dammalapati \textit{et al.}~\cite{Dametal:2009a}}	&0.0&$69(3)$	&$-63(1)$&& $-143(1)\phantom{.}$\\\\
\multicolumn{4}{l}{With the $\delta\langle r^2\rangle$ of:}\\
&Baird \textit{et al.}	\cite{Baietal:79a} 		&0.0&$65.0$		&$\phantom{-}76.5\phantom{()}$&$127.2$&$134.0\phantom{()}$\\
&Bekk \textit{et al.}~\cite{Beketal:79a}		&&\underline{$59.1$}&$\phantom{-}61.7\phantom{()}$&$105.8$&$103.0\phantom{()}$\\
&Sherra \textit{et al.}~\cite{Sheetal:82a}	 	&&$68.7$			&$\phantom{-}63.9\phantom{()}$&$112.5$&$101.5\phantom{()}$\\
&Grundevik \textit{et al.}~\cite{Gruetal:82a}		&&\underline{$59.1$}&$\phantom{-}62.5\phantom{()}$&$106.6$&$105.2\phantom{()}$\\
&van Wijngaarden \textit{et al.}	~\cite{WijLi:95a}&&$51.8$			&\phantom{-}\underline{$56.6$}\phantom{()}&$95.5$	
\end{tabular}\end{ruledtabular}
\end{table*}
Table \ref{tab:valeur_Dam} brings much interesting information.  Most of our values are in complete contradiction with the results of Dammalapati \textit{et al.} The pair $^{138,137}$Ba is consistent but only for the $~^3D_2-~^1P^o_1$ transition. It is true that as regards to the proximity of the two levels $6s5d ~^3D_{1,2}$, it is rather strange that the IS $\delta\nu_{\rm IS}^{138,137}$=114~MHz for one transition is reduced by a factor 2 for the second one.  All the $\delta\nu_{\rm IS}^{138,136}$ and $\delta\nu_{\rm IS}^{138,134}$ values show signs opposite to ours and that, whatever the $\delta\langle r^2\rangle^{138,A}$ used. However, $\delta\nu_{\rm IS}^{138,136}$ are consistent on the absolute scale. As for the $\delta\nu_{\rm IS}^{138,134}$, its determination suffers from our incapacity to discriminate between right and wrong $\delta\langle r^2\rangle^{138,134}$ values.

\begin{table*}
\caption{Extraction of the $\Delta \widetilde K_{\rm MS}$ that would be necessary to reproduce the experimental results of Dammalapati \textit{et al.}~\cite{Dametal:2009a}, combining our theoretical $F$ factor with the $\delta \langle r^2\rangle$ and nuclear masses from the literature. \\$\Delta \widetilde K_{\rm MS}=(\delta\nu_{\rm IS}-F\delta\langle r^2\rangle)/(1/M-1/M')$. \label{tab:Isol_K_ms}}\begin{ruledtabular}
\begin{tabular}{lrcccccc}
	&&	IS$_{\rm exp}$	&$F$ 	&$\delta \langle r^2\rangle$  &$(1/M-1/M')$	&$\Delta \widetilde K_{\rm MS}$\\
	&& (MHz)	& (GHz/fm$^2$)	& (fm$^2$) &	(u$^{-1}$)	&(GHz~u)\\
\hline
\multicolumn{2}{l}{$6s5d~^3D_1-6s6p~^1P^o_1$}\\\\
&	$^{138-137}$Ba&\phantom{$-$}114(4)\phantom{.3.3}
							&0.76&	0.055&	$-0.000053\rightarrow$&$-1366.4$\\
&$^{138-136}$Ba&	$\phantom{1}-59.3(6)\phantom{.3}$
							&0.76&	0.035&	$-0.000107\rightarrow$	&\phantom{$-1$}803.4\\
&$^{138-135}$Ba&\phantom{$-1$}39(4)\phantom{.3.3}
							&0.76&	0.072&	$-0.000161\rightarrow$	&\phantom{$-12$}96.1\\
&$^{138-134}$Ba&\phantom{$-$}114.1(1.0)
							&0.76&	0.040&	$-0.000217\rightarrow$	&\phantom{1}$-386.7$\\
&$^{136-134}$Ba&$-84.8(8)$		&0.76	&0.005&	$-0.000110\rightarrow$	&\phantom{$-1$}805.6\\
&$^{137-135}$Ba&$-75.3(5)$		&0.76&0.017&$-0.000108\rightarrow$&\phantom{$-1$}813.8\\
\hline
\multicolumn{2}{l}{$6s5d~^3D_2-6s6p~^1P^o_1$}\\\\
&$^{138-137}$Ba	&\phantom{$-$}\phantom{1}69(3)	
								&0.74	&0.055&	$-0.000053\rightarrow$	&	$-536.9$\\
&$^{138-136}$Ba&$-\phantom{1}63(1)$	&0.74	&0.035&	$-0.000107\rightarrow$	&\phantom{$-$}831.6\\
&$^{138-135}$Ba&/			\\
&$^{138-134}$Ba&$-143(1)$			&0.74	&0.040&	$-0.000217\rightarrow$	&\phantom{$-$}795.8\\
&$^{136-134}$Ba&$-\phantom{1}80(1)$	&0.74&0.005&$-0.000110\rightarrow$&\phantom{$-$}761.1\\
\end{tabular}
\end{ruledtabular}
\end{table*}

In order to circumvent the problem, table~\ref{tab:Isol_K_ms} presents  the $\Delta K_{\rm MS}$ values that would  reproduce the experimental results of Dammalapati \textit{et al.}~\cite{Dametal:2009a}. In order to do so, we trusted the values of our $F$ factors and adopted the most trustable $\delta \langle r^2\rangle$ and the nuclear masses from the literature. This is done with the help of the formula $\Delta K_{\rm MS}$= (IS$-F\delta \langle r^2 \rangle$)/$(1/M_{138}-1/M')$. For the comparison, we deduced a value for $\delta \langle r^2 \rangle^{136,134}$ by making the difference $\delta \langle r^2 \rangle^{136,134}=\delta \langle r^2 \rangle^{138,134}-\delta \langle r^2 \rangle^{138,136}=0.005$~fm$^2$, as well as for $\delta \langle r^2 \rangle^{137,135}=\delta \langle r^2 \rangle^{138,135}-\delta \langle r^2 \rangle^{138,137}=0.017$~fm$^2$.

In theory, the $\Delta K_{\rm MS}$ parameter is isotope-independent and its value should be identical for a given transition. Table \ref{tab:Isol_K_ms} clearly reveals the incompatibility of the $\Delta K_{\rm MS}$ values within a given transition.

\section*{Conclusion}
The analysis of the $6s^2~^1S_0-6s6p~^{3,1}P^o_1$ and $6s^2~^1S_0-6p^2~^3P_0$ transitions, gave us confidence in our calculations of the electronic factors. 
At the end of their paper, Dammalapati \textit{et al.}~\cite{Dametal:2009a}  suggest that ``the nuclear spin gives rise to an additional contribution to the IS'' for odd isotopes, but on the basis of our new results $-$ and also the presence of discrepancies for even isotopes $-$ their statement is open to doubt.
However, 
complementary investigations would be highly valuable. For instance, with respect to the large MS effects found for these transitions,  further investigation of the convergence of the \textit{ab initio} parameters would be welcome to confirm our predictions. 
The convergence of the electronic parameters with respect to more elaborate correlation models should be investigated 
to refine the estimation of the accuracy of our electronic parameters. Unfortunately, the present calculations have reached the limits of the current computational limits.
Furthermore, by studying other transitions, more confidence could be obtained on the nuclear mean-square radii between two isotopes. 
To continue this study of the nuclear mean-squares charge radii of barium, many investigations remain possible. 
One of the other possibility would be to reinvestigate the Ba II system, that also presents many experimental studies \cite{Hovetal:82a, Wendetal:88a} and only few relativistic calculations \cite{Fisetal:74a}. Another interesting track would be to (re)investigate experimentally isotope shifts of transitions involving the $6s5d~^3D_J$ levels.

\begin{acknowledgments}
This work was supported by the Communaut\'e fran\c caise of Belgium (Action de Recherche Concert\'ee), the Belgian National Fund for Scientific Research (FRFC/IISN Convention) and by the IUAP - Belgian State Science Policy (BriX network P7/12). C. Naz\'e is grateful to the ``Fonds pour la formation \'a la Recherche dans l'Industrie et dans l'Agriculture'' of Belgium for a Ph.D. grant (Boursier F.R.S.-FNRS).
\end{acknowledgments}
\newpage
%

\end{document}